\documentclass{vldb}

\usepackage[T1]{fontenc}
\usepackage{xspace}
\usepackage{url}
\usepackage{epsfig}
\usepackage{color}
\usepackage{times}
\usepackage{pgf}
\usepackage{tikz}
\usepackage{pgfplots}
\usepackage{pgfplotstable}
\usepackage{alltt}
\usepackage{booktabs}
\usepackage{tabularx}
\usepackage{enumitem}
\usepackage{comment}
\usepackage{textgreek}
\usepackage{paralist}
\usepackage[textfont=bf,labelfont=bf]{caption}
\usepackage{subcaption}
\usepackage{etoolbox}

\makeatletter
\def\@copyrightspace{\relax}
\makeatother

\renewenvironment{itemize}{
   \begin{list}{\labelitemi}{
     \setlength{\topsep}{0.5ex}
     \setlength{\itemsep}{-0pt}
     \setlength{\itemindent}{0pt}
     \setlength{\leftmargin}{\labelwidth}
     \addtolength{\leftmargin}{-8pt}}
}{\end{list}}

\newcommand{\firstTxnState}{\textbf{Unprocessed}}
\newcommand{\secondTxnState}{\textbf{Executing}}
\newcommand{\thirdTxnState}{\textbf{Complete}}

\newcommand{\systemName}{\textsc{Bohm}}

\newcommand{\noeditingmarks}{}

\newcommand{\textred}[1]{\textcolor{red}{#1}}

\ifx\noeditingmarks\undefined
   \newcommand{\pgwrapper}[2]{\textbf{#1: }\textred{\textit{#2}}}
   
\else
   \newcommand{\pgwrapper}[2]{}
   
\fi

\usepackage{soul}

\begin{document}

\twocolumn
\newpage
\setlength{\belowcaptionskip}{-12pt}
\title{
  \vspace{-1em} 
  Rethinking serializable multiversion concurrency control\\
(Extended Version) \\
}

\numberofauthors{2}
\author{
\alignauthor
Jose M. Faleiro\\
 \affaddr{Yale University}\\
 \email{jose.faleiro@yale.edu}
 \alignauthor
 Daniel J. Abadi\\
 \affaddr{Yale University}\\
 \email{dna@cs.yale.edu}
 }

\maketitle

\begin{abstract}

Multi-versioned database systems have the potential to significantly increase the amount of concurrency in transaction processing because they can avoid read-write conflicts. Unfortunately, the increase in concurrency usually comes at the cost of transaction serializability. If a database user requests full serializability,  modern multi-versioned systems significantly constrain read-write concurrency among conflicting transactions and employ expensive synchronization patterns in their design. In main-memory multi-core settings, these additional constraints are so burdensome that multi-versioned systems are often \textit{significantly} outperformed by single-version systems.

We propose \systemName{}, a new concurrency control protocol for main-memory multi-versioned database systems. \systemName{} guarantees serializable execution while ensuring that reads \textit{never} block writes. In addition, \systemName{} does not require reads to perform any book-keeping whatsoever, thereby avoiding the overhead of tracking reads via contended writes to shared memory. This leads to excellent scalability and performance in multi-core settings. \systemName{} has all the above characteristics without performing validation based concurrency control. Instead, it is pessimistic, and is therefore not prone to excessive aborts in the presence of contention.  An experimental evaluation shows that \systemName{} performs well in both high contention and low contention settings, and is able to dramatically outperform state-of-the-art multi-versioned systems despite maintaining the full set of serializability guarantees.

\end{abstract}

\section{Introduction}

Database systems must choose between two alternatives for handling
record updates: (1) overwrite the old data with the new data
(``update-in-place systems'') or (2)
write a new copy of the record with the new data, and delete or reduce
the visibility of the old record (``multi-versioned systems''). 
The primary advantage of multi-versioned systems is that transactions
that write to a 
particular record can proceed in parallel with transactions that read
the same record; read transactions do not block write transactions
since they can read older versions until the write transaction has
committed. On the other hand, 
multi-versioned systems must consume
additional space to store the extra versions, and incurs additional
complexity to maintain them. As space becomes increasingly
cheap in modern hardware configurations, the balance is shifting,
and the majority of recently architected database systems
are choosing the multi-versioned approach.

While concurrency control techniques
that guarantee serializability in database
systems that use locking to preclude write-write and read-write
conflicts are well understood, it is much harder to guarantee serializability in
multi-versioned systems that enable reads and writes of the same
record to occur concurrently. One popular option that achieves a level
of isolation very close to full serializability is ``snapshot
isolation'' \cite{berenson95critique}. Snapshot isolation guarantees that each transaction, $T$, reads
the database state resulting from all transactions that committed
before $T$ began, while also guaranteeing that $T$ is isolated 
from updates produced by transactions that run concurrently with $T$. 
Snapshot isolation comes very close to fully guaranteeing serializability, and
indeed, highly successful commercial database systems (such as older
versions of Oracle)
implement snapshot isolation
when the user requests the ``serializable'' isolation level
\cite{oracle7}. However, snapshot isolation is vulnerable to serializability
violations \cite{berenson95critique,fekete04read-only}.
For instance, the famous  write-skew anomaly
can occur when two
transactions have an overlapping read-set and disjoint write-set,
where the write-set (of each transaction) includes elements from the shared
read-set \cite{berenson95critique}. Processing such transactions using snapshot isolation can
result in a final state that cannot be produced if the transactions
are processed serially.

There has been a significant amount of work on making multi-versioned
systems serializable, either by avoiding the write-skew anomaly in
snapshot isolation systems \cite{allocating-isolation,fekete05making},
or by using alternative concurrency control protocols to snapshot
isolation \cite{cahill08serializable,larson11concurrency}. However,
these solutions either severely restrict concurrency in the presence
of read-write conflicts (to the extent that they offer almost no
additional \textit{logical} concurrency as compared to
single-versioned systems) or they require more coordination and
book-keeping, which results in poorer performance 
in main-memory multi-core settings (Section~\ref{sec:mot}). 

In this paper, we start from scratch, and propose \systemName{}, 
a new concurrency control protocol for multi-versioned database
systems. 
The key insight behind \systemName{} is that the complexity of determining a valid
  serialization order of transactions can be eliminated by 
separating concurrency control and version management from transaction
execution. 
Accordingly, \systemName{} determines the serialization order of transactions and
creates versions corresponding to transactions's writes \textit{prior}
to their execution (Section~\ref{sec:design}).
As a consequence of this design, \systemName{}, guarantees full serializability while ensuring
that reads \textit{never} block writes. Furthermore, \systemName{} does not
require the additional coordination and book-keeping introduced by
other methods for achieving serializability in multi-versioned
systems.  The final result is perhaps the most scalable (across
multiple cores) concurrency control protocol ever proposed --- there
is no centralized lock manager, almost all data structures are
thread-local, no coordination needs to occur across threads except at
the end of a large batch of transactions, and the need for latching or
any kind of atomic instructions is therefore minimized (Section~\ref{sec:ccontrol}). 

The main disadvantage of our approach is that entire transactions
must be submitted to the database system before the system can begin
to process them. Traditional cursor-oriented database access, where
transactions are submitted to the database in pieces, are therefore not
supported. Furthermore, the write-set of a transaction must be
deducible before the transaction begins --- either through explicit
write-set declaration by the program that submits the transaction, or
through analysis of the transaction by the database system, or through
optimistic techniques that submit a transaction for a trial run to
get an initial guess for its write-set, and abort the transaction if the
trial run resulted in an incorrect prediction \cite{calvin-tods, calvin-experiments}. 

Although these disadvantages (especially the first one) 
change the
model by which a user submits transactions to a database system, an
increasingly large number of
performance sensitive applications already utilize
stored-procedures to submit transactions to database systems in order
to avoid paying round-trip communication costs to the database
server. These applications can leverage our multi-versioned concurrency
control technique without any modifications. 

\systemName{} thus presents a new, interesting alternative in the
space of multi-version concurrency control options --- an extremely
scalable technique, at the cost of requiring entire transactions with
deducible write-sets in advance. Experiments show that \systemName{}
achieves linear scalability up to (at least) 20 million record
accesses per second with transactions being processed over dozens of
cores.

In addition to contributions around multi-versioned serializability
and multi-core scalability, a third important contribution of
\systemName{} is a clean, modular design. Whereas traditional database
systems use a monolithic approach, with the currency control and
transaction processing components of the systems heavily
cross-dependent and intertwined, \systemName{} completely separates
these system components, with entirely separate threads performing
concurrency control and transaction processing. This modular design is
made possible by \systemName{}'s philosophy of planning transaction
execution in advance, so that when control is handed over to the execution
threads, they can proceed without any concern for other concurrently
executing transactions. This architecture greatly improves database
engine code maintainability and reduces database administrator
complexity.

\section{Motivation}
\label{sec:mot}

We now discuss two fundamental issues that limit the performance
of current state-of-the-art multi-version concurrency control
protocols: 
\begin{inparaenum}[\itshape a\upshape)]
\item the use of global counters to obtain timestamps, and
\item the cost of guaranteeing serializable execution
\end{inparaenum}

\subsection{Centralized Timestamps}
\label{sec:global_counter}
When a multi-version database system updates the value of a record,
the update creates a new version of the record. Each record may
have several versions simultaneously associated with
it. Multi-version databases therefore require a way to decide which of
a record's versions -- if any -- are visible to a particular
transaction. In order to determine the record visible to a
transaction, the database associates \textit{timestamps} with every
transaction, and every version of a record. 

Multi-version databases typically use a global counter to obtain
unique timestamps. When a transaction needs a timestamp, the database
atomically increments the value of the counter using a latch or an
atomic \textit{fetch-and-increment} instruction. Using a global
counter to obtain timestamps works well when it is shared among a
small number of physical CPU cores but does not scale to high core counts
\cite{tu13speedy}. 

Note that the use of a global counter to assign transactions their
timestamps is a pervasive design pattern in multi-version
databases. The use of a global counter is not restricted to only
systems which implement serializable isolation; implementations of
weaker isolation levels such as snapshot isolation and read committed
also use global counters
\cite{cahill08serializable,larson11concurrency}. These systems are
thus subject to the scalability restrictions of using a global
counter.

In order to address this bottleneck, \systemName{} assigns a total
order to transactions prior to their execution. Each transaction is
implicitly assigned a timestamp based on its position in the total
order. When a transaction is eventually executed, \systemName{}
ensures that the state of the database is identical to a serial
execution of the transactions as specified by the total order.
Assigning a transaction its timestamp based on its position in the
total order allows \systemName{} to use low-overhead mechanisms for
timestamp assignment. For instance, in our implementation, we utilize
a single thread which scans the total order of transactions
sequentially and assigns transactions their timestamps
(Section~\ref{sec:timestamps}). 

\subsection{Guaranteeing Serializability}
\label{sec:guaranteeing}
Multi-version database systems
can execute transactions with greater concurrency than their single
version counterparts. 
A transaction, $T_r$, which reads record $x$ need not block a concurrent
transaction, $T_w$, which performs a write operation on record $x$.
In order to avoid blocking $T_w$,
$T_r$ can read a version of $x$, $x_{old}$ that exists \textit{prior} to the
version produced by $T_w$'s write, $x_{new}$. More generally, multiversioning allows
transactions with conflicting read and write sets to execute without blocking
each other. Unfortunately, if conflicting transactions are processed without
restraint, the resulting execution may not be serializable. In our example, if
$T_r$ is allowed to read $x_{old}$, then it must be ordered before $T_w$ in the
serialization order.

In the formalism of Adya et al.\
\cite{adya00generalized}, the serialization graph corresponding to the above
execution contains an \textit{anti-dependency edge} from $T_r$ to $T_w$.
In order for an execution of transactions to be serializable, the serialization
graph corresponding to the trace of the execution cannot contain cycles.
If $T_r$ were to write another record $y$, and $T_w$ read $y$ 
(in addition to $T_r$'s read of $x$ and $T_w$'s write of $x$),
then the order of $T_r$ and $T_w$'s operations on record $y$ must be
the same as the order of their operations on record $x$.
In particular, $T_w$ must not read $y_{old}$ (the version of $y$ prior
to $T_r$'s write), otherwise, the serialization graph would contain an
anti-dependency edge from $T_w$ to $T_r$, leading to a cycle in the
serialization graph. 

\begin{figure}
\centering
\includegraphics[clip,scale=0.4]{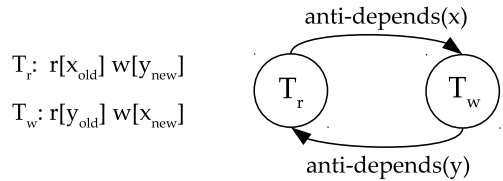}
\caption{ Non-serializable interleaving, and corresponding
  serialization graph of $T_r$ and $T_w$. $r[x_1]$ denotes to a \textit{read} of version 1 of record $x$, correspondingly,
$w[x_1]$ denotes a \textit{write} to record $x$, which produces version 1.
A record's subscript corresponds to the version read or written by the
transaction.
}
\label{fig:interleaving}
\end{figure}

Figure~\ref{fig:interleaving} shows the the interleaved execution of
transactions $T_r$ and $T_w$, and the corresponding serialization
graph. The graph contains two anti-dependency edges,
one from $T_r$ to $T_w$, and the other from $T_w$ to $T_r$; these two edges
form a cycle, implying that the interleaving of $T_w$ and $T_r$ as described
above is not serializable. This example is a variant of Snapshot Isolation's
well known \textit{write-skew} anomaly \cite{berenson95critique}.

In order to avoid non-serializable executions such as the one described above,
multi-versioned database systems need to account for anti-dependencies among
transactions whose read and write sets conflict. There exist two
ways of accounting for anti-dependencies:

\begin{itemize}

\item \textbf{Track Reads.} Whenever a transaction reads a record, the
system tracks the fact that the transaction performed the read by
associating some meta-data with each record in the database. The read
meta-data associated with records in the database system is then used
to decide on the order of transactions.  For instance, the pessimistic
version of Hekaton's multi-version concurrency control algorithm
associates a counter with every record in the database
\cite{larson11concurrency}. The counter reflects the number of
in-flight transactions that have read the record.  As another example,
Cahill et al.\ modify BerkeleyDB's lock manager to track
anti-dependency edges to and from a particular transaction
\cite{cahill08serializable}.

\item \textbf{Validate Reads.} A transaction locally keeps track of
the version of each record it observed. When the transaction is ready
to commit, it validates that the reads it observed are consistent with
a serial order.  This technique is used by Hekaton's
optimistic concurrency control protocol \cite{larson11concurrency},
and Multi-version General Validation \cite{agrawal87distributed}.

\end{itemize}

While both approaches ensure that all executions are
serializable, they come at a cost. Concurrency control protocols track
reads in order to \textit{constrain} the execution of concurrent
readers and writers. For instance, Hekaton's pessimistic concurrency
control protocol does not allow a writer to commit until all
concurrent readers have either committed or aborted
\cite{larson11concurrency}.  In addition to the reduction in
concurrency resulting from the concurrency control protocol itself, tracking
reads entails writes to shared memory. If a record is popular, then
many different threads may attempt to update the same memory words
concurrently, leading to contention for access to internal data
structures, and subsequent cache coherence slow-downs. Since \textit{reads} are
tracked, this contention is present even if the workload is read-only.

The ``Validate Reads'' approach does not suffer from the problem of
requiring reads to write internal data to shared memory. However, validation
protocols reduce concurrency among readers and writers by aborting
readers. Such a situation runs
counter to the original intention of multi-version concurrency
control, because allowing multiple versions of a record is supposed to
allow for greater concurrency among readers and writers.

In order to address these limitations, 
we designed \systemName{}'s concurrency control protocol with the 
following goals in mind: (1) A transaction, $T_r$, which reads the value of a
particular record should \textit{never} block or abort a concurrent transaction
that writes the same record. This should be true whether or not
$T_r$ is a read-only transaction.
(2) Reading the value of a record should not require any writes to shared
memory. 

\section{Design}
\label{sec:design}

\systemName{}'s design philosophy
is to eliminate or reduce coordination among database threads due to
synchronization based on writes to shared memory. \systemName{} ensures that
threads either make decisions based on local state, or amortize the cost of
coordination across several transactions.
\systemName{} achieves this goal by separating concurrency
control logic from transaction execution logic. This separation is reflected
in \systemName{}'s architecture: a transaction is processed by two
different sets of threads in two phases: (1) a
concurrency control phase which determines the proper serialization
order and creates a data structure that will enable the second phase
to process transactions without concern for concurrently executing
transactions, and (2) an execution phase, which actually executes
transaction's logic.

While the separation of concurrency control logic and transaction execution
logic allows \systemName{} to improve concurrency and avoid scalability bottlenecks, it comes
at the cost of extra requirements. In order to plan
execution correctly, the concurrency
control phase needs advance knowledge of each transaction's write-set.
This requirement is not
unique to \systemName{} --- several prior systems exploit a priori information about
transactions's read- and/or write-sets
\cite{aguilera07sinfonia,faleiro14lazy,calvin-tods,moraru13egalitarian}.
These previous systems have shown that even though they need
transactions' write- (and sometimes also read-) sets in advance, it is not necessary for transactions to pre-declare these
read-/write-sets. For example,
Calvin proposes a speculative technique which predicts each transaction's read-/write-sets
on the fly \cite{calvin-tods}. Furthermore,
Ren et. al. show that aborts due to
speculative read/write-set prediction are rare, since the volatility
of data used to derive the read and write sets is usually low\footnote{For example, on
TPC-C, no aborts due to speculative read/write set prediction are
observed.}~\cite{calvin-experiments}. \systemName{} can make use of this technique if
transactions' write-sets are not available (or derivable) in
advance. However, either way, there is a requirement that the entire transaction
be submitted to the system at once. Thus, cursor-oriented transaction
models that a submit a transaction to the system in pieces cannot be supported.

\subsection{System Overview}
\label{sec:system_overview}
Transactions that enter the system are handed over to a single thread
which creates a log in
shared-memory containing a list of all transactions that have been
input to the system. The position of a transaction in this log is
its timestamp. The log is read (in parallel) by $m$ concurrency
control threads. These threads own a logical partition of the records in the
database. For each transaction in the log, each concurrency control
thread analyzes the write-set of the transaction to see if it will
write any records in the partition owned by that thread. If so, the
thread will create space (a ``placeholder'') for the new version in
the database (the contents remain 
uninitialized) and link it to the placeholder associated with the previous
version of record (which was written by the same thread).

A separate set of $n$ threads (the ``transaction execution
threads'') read
the same log of input transactions and perform the reads associated with the
transactions in the log and fill in the pre-existing allocated space
for any writes that they perform. These transaction execution
threads do not start working on a batch of transactions until the the
concurrency control threads have completed that same batch. Therefore,
it is
guaranteed that placeholders already exist for any writes that these
threads perform. Furthermore, reads can determine which version of a
record is the correct version to read (in order to guarantee
serializability) by navigating the backward
references of the placeholders until the record is reached which was
created by a transaction older than the transaction which is performing
the read and invalidated by a transaction newer than the transaction
which is performing the read. If this placeholder associated the
correct version to read remains
uninitialized, then the read must block until the write is
performed. Hence, in \systemName{}, reads never block writes, but
writes can block reads.

The following two subsections give more details on the concurrency
control phase and the transaction execution phase,
respectively. Furthermore, they explain how our design upholds our
philosophy of not allowing contented writes to shared memory, nor any thread
synchronization at the record or transaction granularity.   

\subsection{Concurrency Control}
\label{sec:ccontrol}
The concurrency control layer is responsible for (1) determining the
serialization order of
transactions, and (2) creating a safe environment in which the execution
phase can run transactions without concern for other transactions
running concurrently.

\subsubsection{Timestamp Assignment}
\label{sec:timestamps}
The first step of the concurrency control layer is to insert each transaction into a log in 
main-memory. This is done by a single thread dedicated solely to this
task. Because the concurrency control layer is separated from
(and run prior to) transaction execution, \systemName{} can use this
log to implicitly assign timestamps to transactions (the timestamp of
a transaction is its position in the log). Since a single thread
creates the log prior to all other steps in transaction processing,
log creation (and thus timestamp assignment) is an uncontended
operation. This distinguishes \systemName{} from other multi-versioned
schemes that assign timestamps (which involve updating a shared
counter) as part of transaction processing. Thus, timestamp assignment is
an example of our design philosophy of avoiding
writing and reading from shared data-structures as much as possible.

Several prior multi-version concurrency control mechanisms assign each
transaction, $T$,
two timestamps, $t_{begin}$ and $t_{end}$
\cite{agrawal87distributed,berenson95critique,cahill08serializable,larson11concurrency}.
$t_{begin}$ determines which versions of pre-existing records are visible to
$T$, while $t_{end}$ 
determines the logical time at which $T$'s writes become visible to
other transactions, and is used to validate whether $T$ can commit.
The time between $t_{begin}$ and $t_{end}$ determines the logical interval of
time during which $T$ executes. If another transaction's logical interval overlaps
with that of $T$, then the database system needs to ensure that the transactions
do not conflict with each other (what exactly constitutes a conflict depends on
the isolation level desired). 

In contrast, \systemName{} assigns each transaction a \textit{single
  timestamp},
$ts$ (determined by the transaction's position in the log).
Intuitively, $ts$ ``squashes'' $t_{begin}$ and $t_{end}$ together;
$ts$ determines both the logical time at which $T$ performs its reads,
and the logical time at which $T$'s writes are visible to other transactions.
As a consequence, each transaction appears to execute atomically at time $ts$.

\subsubsection{Inserting Placeholders}
\label{sec:intra_parallelism}
Once a transaction's timestamp has been determined, the concurrency
control layer inserts a new version for every record in the
transaction's write-set.  This includes creating new
versions for index key-values updated by the transaction.
The version inserted by the concurrency control layer contains a placeholder for
the value of the version, but the value is uninitialized. The actual
value of the version is only produced once the corresponding
transaction's logic is executed by the execution layer
(Section~\ref{sec:execution_layer}).

\begin{figure}
\centering
\includegraphics[clip,width=0.4\linewidth]{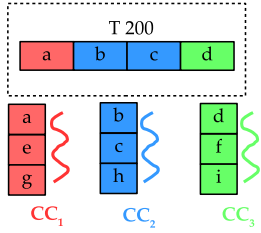}
\caption{Intra-transaction parallelism. Transaction 200, which writes four
records is shown in the upper rectangle. The logical partitioning of
concurrency control thread responsibility is shown below.} 
\label{fig:single_txn}
\end{figure}

Several threads contribute to the processing of a single transaction's
write-set.  \systemName{} partitions the responsibility for each
record of a table across the set of concurrency control threads. When
the concurrency control layer receives a transaction, \textit{every}
concurrency control thread examines $T$'s write-set in order to
determine whether any records belong to the partition for which it is
responsible.

Figure~\ref{fig:single_txn} illustrates how several threads cooperatively
process each transaction. The transaction is assigned a timestamp of 200, and
its write-set consists of records $a$, $b$, $c$, and $d$.
The concurrency control layer partitions records among three threads, $CC_1$,
$CC_2$, and $CC_3$.
$CC_1$'s partition contains record $a$, $CC_2$'s partition
contains records $b$ and $c$, and $CC_3$'s partition contains record $d$.
$CC_1$ thus inserts a new version for record $a$, $CC_2$ does the same for
records $b$ and $c$, and $CC_3$ for $d$. \systemName{} uses
several threads to process a \textit{single} transaction, a form of 
\textit{intra-transaction} parallelism.

Every concurrency control thread must check whether a transaction's
write-set contains records that belong to its partition. For instance,
if record $d$ belonged to $CC_1$'s partition instead of $CC_3$'s,
$CC_3$ would still have to check the transaction's write-set in order
to determine that no records in the transaction's write-set map to its
partition.

This design is consistent with our philosophy that concurrency control
threads should \textit{never} need to coordinate with each other in
order to process a transaction.  Each record is \textit{always}
processed by the same thread (as long as the partitioning is not adjusted); two concurrency control threads will
never try to process the same record, even across transaction
boundaries. The decision of which records of a transaction's write-set
to process is a purely \textit{thread local} decision; a concurrency
control thread will process a particular record only if the record's
key resides in its partition.

Not only does this lead to reduced cache coherence traffic, but it
also leads to multi-core scalability. As we dedicate more concurrency
control threads to processing transactions, throughput increases for
two reasons. First, each transaction is processed by a greater number
of concurrency control threads, which leads to an increase in
intra-transaction parallelism. Since concurrency control threads
do not need to coordinate with each other, there is little downside to
adding additional threads as long as there are enough processing resources
on which they can run. Second, for a fixed database size, the number
of keys assigned to each thread's partition \textit{decreases}.  As a
consequence, each concurrency control thread will have a smaller cache
footprint.

One impediment to scalability is the fact that every concurrency
control thread must examine every transaction that enters the
system. This is logic which is effectively executed serially, since
every concurrency control thread runs the same piece of logic.
Increasing the number of concurrency control threads beyond a certain
point will therefore yield a diminishing increase in throughput due to
Amdahl's law.  Although we have not encountered this scalability
bottleneck in our experimental evaluation, a straightforward mechanism
around the issue is to pre-process transactions prior to handing them
over to the concurrency control layer. The pre-processing layer can
analyze each transaction to determine the set of concurrency control
threads responsible for the writes the transaction performs, and then
forward transaction references to the appropriate concurrency control
threads. Each transaction can be pre-processed independently of
others, thus making the pre-processing step embarrassingly parallelizable.

\begin{figure}
\centering
\includegraphics[clip,width=.8\linewidth]{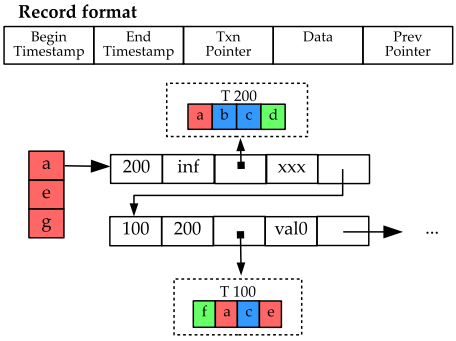}
\caption{Inserting a new version}
\label{fig:insertion}
\end{figure}

\subsubsection{Processing a single transaction's read/write set}
\label{sec:placeholders}
For each record in a transaction's {\bf write-set}, the concurrency control phase produces a new
version to hold the transaction's write.
Figure~\ref{fig:insertion} shows the format of a record version.
Each version consists of the following fields:

\begin{itemize}
\item \textbf{Begin Timestamp.} The timestamp of the transaction that created
  the record.

\item \textbf{End Timestamp.} The timestamp of the transaction that invalidated
  the record.

\item \textbf{Txn Pointer.} A reference to the transaction
  that must be executed in order to obtain the value of the record.

\item \textbf{Data.} The actual value of the record. 

\item \textbf{Prev Pointer.} A reference to the version of the record that
  precedes the current version.
  
\end{itemize}

When inserting a new version of a record, the concurrency control thread sets
the version's fields as follows: (1) the version's start timestamp is set to
the timestamp of the transaction that creates the version, (2) the version's
end timestamp is set to infinity, (3) the version's txn pointer is set to the
transaction that creates the version, (4) the version's data is left uninitialized,
(5) the version's prev pointer is set to the preceding version of the record.

Figure~\ref{fig:insertion} shows the thread $CC_1$ inserting a new version of
record $a$, which is produced by transaction $T_{200}$. 
$CC_1$ sets the new version's begin timestamp to 200, and its end timestamp to
infinity. The version's txn pointer is set to $T_{200}$ (since $T_{200}$
produces the new version). 
At this point, the version's data has not yet been produced;
\systemName{} needs to execute $T_{200}$ in
order to obtain the value of the version.

While inserting a new version of record $a$, $CC_1$ finds that a previous
version of the record exists. The older version of $a$ was produced by
transaction $T_{100}$. $CC_1$ sets the new version's prev pointer to the old
version, and sets the old version's end timestamp to 200.

In order to create a new version of a record, \systemName{} does not need to
synchronize concurrency control threads. \systemName{} partitions the database
among concurrency control threads such that a record is \textit{always}
processed by the same thread, even across transaction boundaries (Section
\ref{sec:intra_parallelism}).
One consequence of this design is that
there is no contention in the concurrency control phase. The
maintenance of the pointers to the current version of every record can be done in a
thread-local data-structure; thus the look-up needed to populate the
prev pointer in the new versions is thread-local. Furthermore, if multiple
transactions update the same hot record, the corresponding new versions of the
record are written by the \textit{same} concurrency control thread,
thereby avoiding cache coherence overhead.

For each element in the transaction's {\bf read-set}, \systemName{} needs to identify
the corresponding version that the transaction will read. In general,
the concurrency control phase does not need to get involved in
processing a transaction's read-set. When an execution thread that is
processing a transaction with timestamp $ts$ wants to
read a record in the database,
it can find the correct version of the record to read by starting at
the latest version of the record, and following the chain of linked
versions (via the prev pointer field) until
it finds a version whose 
$t_{begin} \le ts$ and $t_{end} \ge ts$. If no such version exists, then the
record is not visible to the transaction. 

While the above-described technique to find which version of a
record to read is correct, the cost of the traversal of pointers
may be non-trivial if
the linked list of versions is long. Such a situation may arise if a record
is popular and updated often. An optimization to eliminate
this cost is possible if the concurrency control phase has advanced
knowledge of the read-sets of transactions (in addition to the
write-set knowledge it already requires). In this case, for every
record a transaction will read, concurrency control threads annotate
the transaction with a reference to the correct version of the record
to read. This is a low-cost operation for the concurrency control
threads since the correct version is simply the most recent version at the time the
concurrency control thread is running\footnote{This is true since concurrency control threads
process transactions sequentially (threads derive concurrency by exploiting
intra-transaction parallelism).}.
In particular, if a record in a transaction's read-set resides on a
concurrency control thread's logical partition, the thread looks up
the latest version of the record and writes a reference to the latest
version in a memory word reserved in advance within the
transaction. The concurrency control thread \textit{does not} track
the read in the database, it merely gives the transaction a reference
to the latest version of the record as of the transaction's timestamp.

A consequence of \systemName{}'s design is that a transaction's reads
do not require any contended writes to shared memory. Even for
the read-set optimization mentioned above, the write containing the
correct version reference for a read is to pre-allocated space for the
reference within a transaction, and is uncontended since only one
concurrency control thread is responsible for a particular record.
In contrast, pessimistic multi-version systems such as Hekaton
\cite{larson11concurrency} and Serializable Snapshot Isolation
\cite{cahill08serializable} need to coordinate a transaction's reads
with concurrent transaction's writes in order to avoid 
serializability violations.

\subsubsection{Batching}
Only after a transaction $T$ has been processed by all appropriate concurrency
control threads can it be handed off to the transaction execution layer.
One na\"{\i}ve way of performing this hand-off is for the concurrency
control threads to notify each other after having processed each
transaction by using synchronization barriers.
After processing $T$, each concurrency control thread enters a global barrier in
order to wait for all other threads to finish processing $T$. After
all threads have entered this barrier, each concurrency
control thread can begin processing the next transaction.

Unfortunately, processing transactions in this fashion is extremely
inefficient.  Threads need to synchronize with each other on every
transaction, which has the effect of forcing concurrency control
threads to effectively execute in lock step.  Another issue is that
some concurrency control threads are needlessly involved in this
synchronization process. Consider a scenario where none of the records
in $T$'s write-set belong to a concurrency control thread, $CC$'s,
partition.  $CC$ has to wait for every thread in order to move on to
the next transaction despite the fact that $CC$ ``contributes''
nothing to $T$'s processing.

\systemName{} avoids expensive global coordination on every
transaction, and instead \textit{amortizes} the cost of coordination
across large batches of transactions. The concurrency control thread
responsible for allotting each transaction a timestamp accumulates
transactions in a batch.  The concurrency control threads responsible
for writing versions receive an ordered batch of transactions, $b$, as
input. Each concurrency control thread processes every transaction in
$b$ independently, without coordinating with other threads
(Sections \ref{sec:intra_parallelism}, \ref{sec:placeholders}).  Once
a thread has finished processing every transaction in $b$, it enters a
global barrier, where it waits until all concurrency control threads
have finished processing $b$, amortizing the cost of a single global
barrier across every transaction in $b$.

Coordinating at the granularity of batches means that some threads may
outpace others in processing a batch; a particular thread could be
processing the $100^{th}$ transaction in the batch while another is
still processing the $50^{th}$ transaction. Allowing certain
concurrency control threads to outpace others is safe for the same
reason that intra-transaction parallelism is safe
(Section~\ref{sec:intra_parallelism}): \systemName{} partitions the
database among concurrency control threads such that a particular record is
\textit{always} processed by the same thread, even across transaction
boundaries.

\subsection{Transaction Execution}
\label{sec:execution_layer}
After having gone through the concurrency control phase, a batch of transactions
is handed to the transaction execution layer.
The execution layer performs two main functions: it executes transactions'
logic, and (optionally) incrementally garbage collects versions which are no longer visible
due to more recent updates.

\subsubsection{Evaluating Transaction Logic}
\label{sec:txn_logic}
The concurrency control layer inserts a new version for every record
in a transaction's write-set.  However, the \textit{data} within the
version cannot yet be read because the transaction responsible for
producing the data has not yet executed; concurrency control threads
merely insert placeholders for the data within each record.  Each
version inserted by the concurrency control layer contains a reference
to the transaction that needs to be evaluated in order to obtain the
data of the version.

\textbf{Read Dependencies.}
Consider a transaction $T$, whose read-set consists of ${r_1, r_2, ..., r_n}$.
$T$ needs to read the correct version of each record in its read-set
using the process described in Section~\ref{sec:placeholders}. 
However, the data stored inside one or more of these correct versions may not yet
have been produced because the corresponding transaction has not yet been
executed. Therefore, an execution thread may not be able to complete
the execution of $T$ until the transaction upon which $T$ depends has
finished executing.

\textbf{Write Dependencies.}
Every time the value of a particular record is updated, the
concurrency control layer
creates a new version of the record, stored separately from other versions. 
Consider two transactions
$T_1$ and $T_2$, such that (1) neither transactions' logic
contain aborts, and (2) $T_1$ is processed before $T_2$ by the concurrency
control layer. Both transactions' 
write-sets consist of a single record, $x$, while their read-sets do not contain
record $x$. 
In this scenario, the concurrency control layer will write out two versions
corresponding to record $x$, one each for $T_1$'s and $T_2$'s update.
The order of both transaction's updates is already decided by the concurrency
control layer; therefore, $T_1$ and $T_2$'s execution need not be coordinated.
In fact, $T_2$ could execute \textit{before} $T_1$, despite the fact that
$T_1$ precedes $T_2$, and their write-sets overlap. However, 
if $T_2$ performs a read-modify-write of record $x$, 
then $T_2$  must wait for the version of $x$ produced by $T_1$ before it can
proceed with the write (this is a type of read dependency explained above).
If $T_2$ aborts, then it also needs to wait for $T_1$. The
reason is that in this case, the data written to its version of $x$
is equal to that produced by $T_1$. Thus, $T_2$ has a read dependency
on $T_1$. 

We now describe how a set of execution threads execute a batch of transactions
handed over from the concurrency control layer. 
The execution layer receives a batch of transactions in an ordered array
$<T_0, T_1, ..., T_n>$.
The transactions are partitioned among $k$ execution threads such that
thread $i$ is responsible for ensuring transactions $T_i$, $T_{i+k}$, $T_{i+2k}$, and so
forth are processed. Thread $i$ does not need to directly execute all
transactions that it is responsible for --- other threads are
allowed to execute transactions assigned to $i$, and $i$ is allowed to
execute transactions assigned to other threads. 
However, before moving onto a new batch of transactions, thread $i$
must ensure that all transactions that it is responsible for in the current batch have been
executed. 

Each transaction can be in one of three states: \firstTxnState{},
\secondTxnState{}, and \thirdTxnState{}. All transactions received from the
concurrency control layer are in state \firstTxnState{} --- this state corresponds
to transactions whose logic has not yet been evaluated.
A transaction is in state \secondTxnState{} if an execution thread is in the
process of evaluating the transaction.
A transaction whose logic has been evaluated is in state \thirdTxnState{}.

In order to process a transaction, $T$, an execution thread, $E$, attempts to
atomically change $T$'s state from \firstTxnState{} to \secondTxnState{}. 
$E$'s attempt fails if $T$ is either
already in state \secondTxnState{} or \thirdTxnState{}.
If $E$'s attempt is successful, then \systemName{} can be sure that $E$ has exclusive
access to $T$; subsequent transactions that try to change $T$'s state from
\firstTxnState{} to \secondTxnState{} will fail.

If, upon trying to read a record, $E$ discovers a read dependency on a
version that has yet to be produced, $E$ tries to
recursively evaluate the transaction $T'$ which must be evaluated to produce
the needed version. If $E$ cannot evaluate $T'$ (because another thread is already
processing $T'$) then $E$ sets $T$'s state back to \firstTxnState{}.
$T$ is later picked up by an
execution thread (not necessarily $E$) which attempts once again to
execute the transaction. After completing
all reads and writes for $T$, $E$ sets $T$'s state to
\thirdTxnState{}.

Note that execution and concurrency control threads operate on
different batches concurrently.  Execution threads are responsible for
producing the \textit{data} associated with versions written in a
batch, while concurrency control threads create versions and update
the appropriate indexes.  Logically, a version's data is a
\textit{field} associated with the version
(Section~\ref{sec:placeholders},
Figure~\ref{fig:insertion}). Execution threads only write a version's
data field; therefore, there are no write-write conflicts between
execution and concurrency control threads. However, in order to locate
the record whose data must be read or written, execution threads may
need to read database indexes. Execution threads need only coordinate with a
single writer thread while reading an index -- the concurrency control
thread responsible for updating the index entry for that record. 
\systemName{} uses standard latch-free hash-tables to index
data; readers need only spin on inconsistent or stale data
\cite{jung13scalable}. We believe that coordinating structural
modifications (SMOs) by a single writer with multiple readers is significantly
less complex than coordinating multiple writers and readers.
We leave the broader discussion of SMOs in general indexing
structures to future work.

\subsubsection{Garbage Collection}
\label{sec:gc}
\systemName{} can be optionally configured to automatically garbage
collect all versions that are no longer visible to any active or future
transactions. Records that have been ``garbage collected'' can be
either deleted or archived. This section describes how \systemName{} decides when a version can be safely
garbage collected.

 Consider a scenario where a transaction $T$ updates a record $r$, whose
 version preceeding $T$'s update is $v_1$. $T$ produces a new version $v_2$.
 If there exists an unexecuted transaction, $T'$, whose timestamp precedes
 that of $T$, then the version of $r$ visible to $T'$ is $v_1$. We need to keep
 version $v_1$ until all transactions that read version $v_1$ have finished
 executing. This intuition leads to the following general condition for garbage
 collecting old versions of records:
 
 \textbf{Condition 1:} 
 Whenever a transaction updates the value of a particular record, we can garbage
 collect the preceding version of the record when all transactions that read the
 preceding version have finished executing.
 
 While \textbf{Condition 1} is correct, it requires \systemName{} to track every
 transaction that reads each version of a record. However, maintaining this
 meta-data goes
 against \systemName{}'s design philosophy of avoiding all writes to shared
 memory on a read.
 Instead of waiting for precisely the set of transactions
 that read a particular version to complete, we can wait for all transactions
 that precede $T$ to finish executing. This set of transactions is a super-set
 of the transactions that read the preceding version of record
 $r$. Therefore, it is more conservative, but still correct. Instead of tracking every transaction that reads the value of
 a particular version, we now need to only track when every transaction that
 precedes $T$ has finished executing. This leads to the following, more
 efficient, general condition for garbage collecting old versions:
 
 \textbf{Condition 2:}
 Whenever a transaction updates the value of a particular record, we can garbage
 collect the preceding version of the record when all transactions with lower
 timestamps have finished executing.
 
 In order to implement garbage collection based on \textbf{Condition 2}, the
 system needs to maintain a global \textit{low-watermark} timestamp.
 The low-watermark corresponds to the timestamp of
 transaction, $T'$, such that all transactions prior to $T'$ have finished
 executing. Maintaining the low-watermark is less expensive than
 maintaining the meta-data required for \textbf{Condition 1}.
 However, 
 the low-watermark is a shared global variable that is subject to updates
 by every execution thread --- potentially as frequently as once per
 transaction --- which can hinder
 \systemName{}'s scalability. 

As an alternative, note that \systemName{}'s execution layer receives transactions in batches.
Transactions are naturally ordered across batches; if batch $b_0$
precedes batch $b_1$, then \textit{every} transaction in $b_0$
precedes \textit{every} transaction in $b_1$.  Assume that a
transaction $T$ belongs to batch $b_0$. $T$ updates the value of
record $r$, whose version is $v_1$, and produces a new version
$v_2$. The timestamp of any transaction in batch $b_i$, where $i \geq
1$, will always exceed $v_2$, $T$'s timestamp.  As a consequence,
version $v_1$, which precedes $v_2$, will \textit{never} be visible to
transactions in batches which occur after $b_0$. Section
\ref{sec:txn_logic} explained that 
execution threads always process batches sequentially; that is, each
thread will not move onto batch $b_{i+1}$ until the transactions it is
assigned in $b_i$ have been executed. Therefore, $v_1$ can be garbage
collected when \textit{every} execution thread has finished
executing batch $b_0$. This condition holds regardless of which batch
$v_1$ was created in. In fact, $v_1$ may even have been created in
$b_0$. This intuition forms the basis of the condition \systemName{}
actually uses to garbage collect old versions:

\textbf{Condition 3:}
Whenever a transaction in batch $b_i$ updates the value of a particular record,
we can garbage collect the preceding version of the record when every execution
thread has finished processing every transaction in batch $b_i$. 

Garbage collection based on \textbf{Condition 3} is amenable to an
efficient and scalable implementation based on read-copy-update (RCU)
\cite{mckenney01read-copyupdate}. The heart of the technique is
maintaining a global low-watermark corresponding to the minimum batch
of transactions processed by \textit{every} execution thread. Each
execution thread $t_i$ maintains a globally visible variable
$batch_i$, which corresponds to the batch most recently executed by
$t_i$. $batch_i$ is only updated by $t_i$. We designate one of the
execution threads, $t_0$, with the responsibility of periodically
updating a global variable $low watermark$ with $min(batch_i)$, for
each $i$.

\subsubsection{Correctness}
We now sketch an argument for why \systemName{}'s concurrency control protocol
guarantees serializable execution of transactions. Given any pair of
transactions $T_0$ and $T_1$ with timestamps $ts_0$ and $ts_1$, \systemName{}
ensures the following invariant on the resulting transaction
serialization graphs\footnote{See Section~\ref{sec:mot} for a
  description of serialization graphs.}:

\textit{If $ts_0 < ts_1$, then the serialization graph
contains no dependencies from $T_1$ to $T_0$. }

Effectively, this invariant
implies that the timestamp order of transactions is equivalent to their
serialization order.
Serialization graphs may contain three kinds of dependencies among transactions:
write-write (ww) dependencies, write-read (wr) dependencies, and
read-write (rw) dependencies \cite{adya00generalized}.
\systemName{} guarantees that the invariant holds for each kind of dependency:

\begin{itemize}

\item \textbf{ww dependencies.}
  In order for a ww dependency to occur from $T_1$ to $T_0$, their write-sets
  must overlap, and $T_1$'s write must precede $T_0$'s write.
  For each record $r$ in the overlapping part of the write-sets, a single concurrency control thread will write out versions of record $r$
  (Section~\ref{sec:intra_parallelism}). In addition, concurrency control threads
  always process transactions in timestamp order. Since $ts_0 < ts_1$, 
the appropriate concurrency control thread 
  thread will always write out $T_0$'s version before $T_1$'s version.
  Therefore, $T_1$'s write can never precede $T_0$'s. Thus there exist no
  ww dependencies from $T_1$ to $T_0$.
  
\item \textbf{wr dependencies.}
  There exists a wr dependency from $T_1$ to $T_0$ if $T_0$ observes the effects
  of a write by $T_1$. A version of a record whose respective begin and end
  timestamps are $t_{begin}$ and $t_{end}$, is visible to $T_0$ if
  $t_{begin} \le ts_0$ and $t_{end} \ge ts_0$.

  When $T_1$ updates a record, \systemName{} changes $t_{end}$ from
  \textit{infinity} to $ts_1$, while $t_{begin}$ is unaffected.
  Prior to $T_1$'s update, $t_{end} > ts_0$ (since $t_{end}$ is
  \textit{infinity}). After $T_1$'s update, $t_{end} > ts_0$ because
  $ts_1 > ts_0$.
$T_1$'s update does not affect
  $T_0$'s visibility of the record because $t_{end} > ts_0$ both before and
  after $T_1$'s update. The argument for inserts and deletes follows along
  similar lines.
  
\item \textbf{rw dependencies.}
  In order for an rw dependency to occur from $T_1$ to $T_0$, $T_1$ must read a
  record $r$ that $T_0$ writes, and $T_0$'s write must occur \textit{after}
  $T_1$'s read. This implies that $T_0$ creates a new version of $r$ such that
  the version's begin timestamp, $t_{begin} > ts_1$. When $T_0$ creates a new
  version of a record, it sets the version's $t_{begin}$ to $ts_0$. Since
  $ts_0 < ts_1$, $t_{begin} < ts_1$; which contradicts the
  requirement that $t_{begin} > ts_1$ in order for $T_1$'s read to precede
  $T_0$'s write.
  
\end{itemize}

\section{Experimental Evaluation}
\label{sec:exps}
\systemName{}'s primary contribution is a
multi-version concurrency control protocol that is able to achieve
serializability at lower cost than previous multi-version concurrency
control protocols. Therefore, the best
comparison points for \systemName{} are other multi-versioned
protocols. We thus compare \systemName{}'s
performance to two state-of-the-art multi-versioned protocols: the
optimistic variant of Hekaton \cite{larson11concurrency}, and
Snapshot Isolation (implemented within our Hekaton codebase) 
\cite{berenson95critique}.
Our Hekaton and Snapshot Isolation
(SI) implementations include support for \textit{commit dependencies},
an optimization that allows a transaction to speculatively read
uncommitted data. In order to keep our codebase simple, 
our Hekaton and SI implementations do not
incrementally garbage collect versions from the database and use a
simple fixed-size array index to access records (\systemName{} and its other
comparison points discussed below 
use dynamic hash-tables). The lack of garbage 
collection does not negatively impact performance; on the contrary,
garbage collection was cited as one of the primary contributors to
Hekaton's poor performance relative to single-versioned systems.
\systemName{} runs with garbage collection enabled; therefore, any
performance gains of \systemName{} over Hekaton and SI that we see in
our experiments are conservative estimates. We would expect an even
larger performance difference had garbage collection of these
baselines been turned on. 

While Hekaton and SI are the main points against which we seek to
compare \systemName{}, our evaluation also includes single-version
baselines.  We compare \systemName{} against state-of-the-art
optimistic concurrency control (OCC) and two-phase locking
(2PL) implementations. Our OCC implementation is a
direct implementation of Silo~\cite{tu13speedy} -- it
validates transactions using decentralized timestamps and
avoids \textit{all} shared-memory writes for records that were only read.
All our optimistic baselines --- single-version OCC, Hekaton, and SI --- are
configured to retry transactions in the event of an abort
induced by concurrency control.

Our 2PL implementation
uses a hash-table to store information about the locks acquired by
transactions. Our locking implementation has three important
properties. 
\begin{inparaenum}[\itshape a\upshape)]
\item {\bf Fine-grained latching.} We use per-bucket latches on the
  lock table to avoid a centralized latch bottleneck.
\item {\bf Deadlock freedom.} We exploit advance knowledge of
transactions's read- and write-sets to acquire locks in lexicographic
order. Acquiring locks in this fashion is guaranteed
to avoid deadlocks. Consequently, our locking implementation does not
require any deadlock detection logic.
\item {\bf No lock table entry allocations.} We exploit advance
knowledge of a transaction's read- and write-sets to allocate a
sufficient number of lock table entries \textit{prior} to submitting
the transaction to the database. The consequence of this design is
that the duration for which locks are held is reduced to the bare
minimum.
\end{inparaenum}

Our experimental evaluation is conducted on a single 40-core machine,
consisting of four 10-core Intel E7-8850 processors and 128GB of memory. Our
operating system is Linux 3.9.2. All experiments are performed in main-memory,
so secondary storage is not utilized for our experiments.

In all our implementations, there is a 1:1 correspondence between
threads and cpu cores; we explicitly explicitly pin long running
threads to cpu cores. Traditional database systems typically assign a
transaction to a single physical thread. If the transaction blocks,
for instance, while waiting for lock acquisition or disk I/O, the
database yields the thread's processor to other threads with
non-blocked transactions.  In order to adequately utilize processing
resources when transactions block, the database ensures that there are
a sufficiently large number of threads running other non-blocked
transactions.  The number of active threads is therefore typically larger than
the number of physical processors.
In contrast, transactions in single node main memory database systems
do not block on I/O. Therefore, some main memory database systems use
non-blocking thread implementations such that when a transaction
blocks for any reason (such as a a failure to acquire a lock), instead
of yielding control to another thread, the thread temporarily stops
working on that transaction and picks up another transaction to
process, eventually returning to the original transaction when it is
no longer blocked \cite{ren13lightweight}. We leverage this approach
in our implementations, so that all baselines we experiment with
do not need to pay thread context switching costs. 

\subsection{Concurrency control scalability}
\label{sec:microbenchmark}
We begin our experimental evaluation by exploring the effect of
the separation concurrency control from transaction execution in
\systemName{} (Section~\ref{sec:design}). 
Recall that concurrency control and transaction
execution are each handled by two separate modules, each of which is
parallelized by a separate group of threads. Both, the number of threads devoted
to concurrency control and the number of threads devoted to transaction
execution are system parameters that can be varied by a system administrator.
We vary both parameters in this experiment. 

\begin{figure}
\centering
\includegraphics[clip,width=\linewidth]{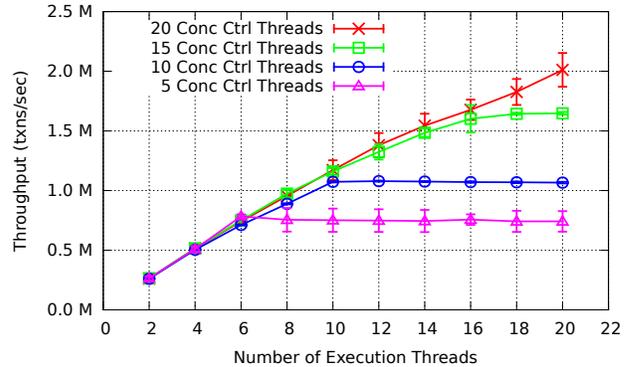}
\caption{Interaction between concurrency control and transaction
  execution modules.}
\label{fig:ccontrol_graph}
\end{figure}

Our experiment stresses the concurrency control layer as
much as possible, in order to test scalability. In particular:
\begin{itemize}

\item  
The workload consists of short, simple transactions, involving only
10 RMWs of different records. Furthermore, each record is very
small (it only contains a single 64-bit integer attribute), and the
modification that occurs in the transaction consists of a simple increment of this integer.
As a consequence, the execution time of each
transaction's logic is very small.

\item
The database consists of 1,000,000 records, and the 10 records
involved in the RMWs of each transaction are chosen from a uniform distribution. As a consequence, transactions
rarely conflict with each other.

\item
  The entire database resides in main memory, so there are no delays to access
  secondary storage.
  
\end{itemize}

As a result of these three characteristics, there are no delays around
contending for data, waiting for storage, or executing transaction
logic. This stresses the concurrency control layer as much as possible
--- it is not able to hide behind other bottlenecks and delays in the
system, and must keep up with the transaction execution layer, which
consumes very little effort in processing each transaction. 

Figure~\ref{fig:ccontrol_graph} shows the results of our
experiment. The number of threads devoted to transaction execution is
varied on the x-axis, while the number of threads devoted to
concurrency control is varied via the 4 separate lines on the graph.  
Recall
that in our experimental setup, there is a 1:1 correspondence between
threads and CPU cores. Thus, adding more threads to either concurrency
control or transaction execution is equivalent to adding more cores
dedicated to these functions.  

Despite the extreme stress on the
concurrency control layer in this microbenchmark, when the number of
concurrency control threads (cores)
significantly outnumber the number of execution threads (cores), the system is
bottlenecked by transaction execution, not concurrency control. 
This is why the throughput
of each configuration initially increases as more execution threads
are added. 
However, once the throughput of the execution layer matches that of
the concurrency control layer, the total throughput plateaus. At this
point, the throughput of the system is bottlenecked by
the concurrency control layer.

As we increase the number of concurrency control threads (represented
by the four separate lines in Figure \ref{fig:ccontrol_graph}), the
maximum throughput of the system increases. This indicates that the
concurrency control layer's throughput scales with increasing thread
(core) counts. This is because the concurrency control layer is able to exploit
greater intra-transaction parallelism and has a lower per-thread cache
footprint at higher core counts (Section~\ref{sec:placeholders}).

While the number of concurrency control and execution threads can be
varied by a system administrator, the choice of the \textit{optimal}
division of threads between the concurrency control and
execution layers is non-trivial. As Figure~\ref{fig:ccontrol_graph}
indicates, using too few concurrency control threads results in under
utilization of execution threads, while using too many
concurrency control threads will constrain overall throughput as not
enough execution threads will be available to process transactions.

This problem can be addressed by using techniques for dynamic load
balancing in high-performance web-servers. 
\systemName{} uses a \textit{staged event-driven architecture} (SEDA)
\cite{welsh01seda}; the concurrency control and execution phases each
correspond to a stage. The processing of a single request (in
\systemName{}'s case, a transaction) is divided between the
concurrency control and execution phases. As advocated by SEDA, there
is a strong separation between the concurrency control and
execution phases; threads in the concurrency control phase are unaware
of threads in the execution phase (and vice-versa). 
SEDA's design allows for dynamic allocation of threads to
stages based on load. By following SEDA's design principles, 
\systemName{} can similarly dynamically allocate
resources to the concurrency control and execution phases. 

Overall, this initial experiment provides evidence of the scalability
of \systemName{}'s design. As we increase the number of concurrency
control and execution threads in unison, the overall throughput scales
linearly. At its peak in this experiment, \systemName{}'s concurrency
control layer is able to handle nearly 2 million transactions a second
(which is nearly 20 million RMW operations per second) --- a number
that (to the best of our knowledge) surpasses any known real-world
transactional workload that exists today.

\subsection{YCSB}
\label{sec:ycsb}
We now compare \systemName{}'s throughput against the implemented
baselines of Hekaton, Snapshot Isolation (SI), OCC, and locking on the
Yahoo! Cloud Serving Benchmark (YCSB) \cite{cooper10ycsb}.

For this set of experiments, we use a single table consisting of
1,000,000 records, each of size 1,000 bytes (the standard record size in YCSB).
We use three kinds of transactions: the first performs 10
read-modify-writes (RMWs) --- just like the experiment
above, the second performs 2 RMWs and 8 reads (which we call 2RMW-8R),
and the third is a read-only transaction which reads 10,000 records.

We use a workload consisting of only 10RMW transactions to compare the
overhead of multiversioning in \systemName{} compared to a single
versioned system. If a workload consists of transactions that perform
only RMW operations, we do not expect to obtain any benefits from
multiversioning. To understand why, consider two transactions $T_1$
and $T_2$ whose read- and write-sets consist of a single record,
$x$. Since both transactions perform an RMW on $x$, their execution
must be serialized. Either $T_1$ will observe $T_2$'s write or
vice-versa.  This serialization is equivalent to how a single version
system would handle such a conflict.

In contrast, we expect that, under high contention, multi-versioned
systems will execute a workload of 2RMW-8R transactions with
greater concurrency than single-versioned systems. The reason is that
if a transaction, $T$, only reads the value of record $r$, then $T$
does not need to block a transaction $T'$ which writes $r$ (or
alternatively, performs an RMW operation on $r$).

Finally, we use a workload consisting of a combination of 10RMW and
read-only transactions to demonstrate the impact of long running
read-only transactions on each of our baselines. We expect such a
workload to favor multi-version systems because multi-version systems
ensure that read-only transactions execute without blocking
conflicting update transactions (and vice-versa).

\subsubsection{10RMW Workload}
\label{sec:ycsb_10rmw}

\begin{figure}
\centering
\includegraphics[clip,width=\linewidth]{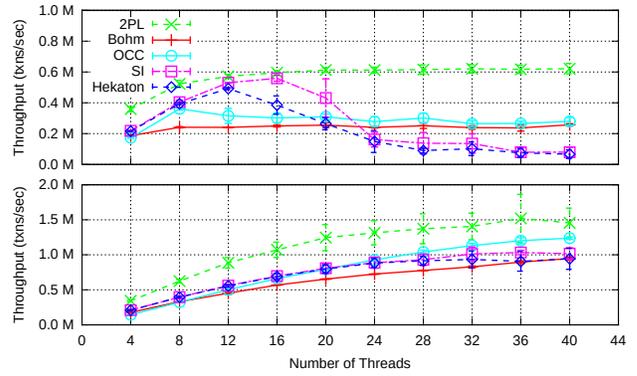}
\caption{YCSB 10RMW throughput. Top: High Contention (theta =
  0.9). Bottom: Low Contention (theta = 0.0).}
\label{fig:ycsb_10rmw}
\end{figure}
Our first experiment compares the throughput of each system on YCSB
transactions which perform 10RMW operations, where each element of a
transaction's read- and write-set is unique. We run the experiment
under both low and high contention. We use a zipfian distribution to
generate the elements in a transaction's read- and write-sets. We vary
the contention in the workload by changing the value of the zipfian
parameter theta \cite{gray94billion-record}. The low contention
experiment sets theta to 0, while the high contention experiments sets
theta to 0.9.

The graph at the top of Figure~\ref{fig:ycsb_10rmw} shows the result
of our experiment under high contention.  
The throughput of every system does not scale beyond a
certain threshold due to the high contention in
the workload --- there are simply not enough transactions that do not
conflict that can be run in parallel. Hekaton and SI
perform particularly poorly when there are a large
number of concurrently executing threads because under high
contention, they are prone to large numbers of aborts. 
 Optimistic systems run transactions
concurrently, regardless of the presence of conflicts, and validate
that transactions executed in a serializable fashion (or in the case
of SI, that write-write conflicts are absent and that transactions
read a consistent snapshot of the database).  A transaction is aborted
if its validation step fails, and the work performed by the
transaction is effectively wasted. 
Note, however, that while OCC is also
optimistic and suffers from aborts; it does not suffer from the
same drop in throughput as
Hekaton and SI. This is because Silo (the version of OCC we use in
these experiments) uses a back-off scheme to slow
down threads when there is high write-write contention.

The reason why \systemName{} is outperformed by the locking
implementation is that individual transactions are subject to greater
overhead. When a multi-version database
system performs an RMW operation on a particular record (say, $x$),
the corresponding execution thread must bring the memory words
corresponding to $x$'s version being read into cache, and write a
\textit{different} set of words corresponding to the new
version of $x$. In contrast, when a single-version system performs an
RMW operation, it writes to the same set of memory words it reads. Note
that the overhead of creating new versions must be paid during a 
transaction's \textit{contention period} \footnote{We define a
transaction's \textit{contention period} as the time period during
which concurrently running conflicting transactions must either block
or abort (depending on the pessimistic or optimistic nature of the
concurrency control protocol).}.
As a
consequence, version creation has a greater negative impact on
throughput in high contention workloads (as compared to low contention
workloads). 
This effect is magnified for YCSB, since the size of each YCSB record
is fairly large (1,000 bytes), and each transaction must therefore pay
the overhead of writing ten new 1,000-byte records. Thus, all the
multi-versioned systems (including \systemName{}) have a disadvantage
on this workload --- they pay the overhead of multi-versioning without
getting any benefit of increase in concurrency for this 100\% RMW
benchmark.

Nonetheless, \systemName{}'s lack of aborts allow it to achieve
over twice the throughput of the other multi-versioned systems
(Hekaton and SI) when there are large numbers of concurrently running threads. 
However, when there are low numbers of concurrently running threads,
there is less contention and the optimistic systems do not suffer from
many aborts. Furthermore, the high theta increases the number of
versions created and ultimately garbage collected for the ``hot''
records, and our configuration of Hekaton and SI to not have to
garbage collect give them a small advantage over \systemName{}.
 
We find that OCC's
throughput begins to degrade between 8 and 12 threads, while Hekaton
and SI are able to sustain higher throughput for slightly higher
thread counts (12 and 16 threads respectively). The reason for this is
that Hekaton and SI use an optimization that allows transactions to
speculatively read uncommitted values (commit dependencies)
\cite{larson11concurrency}. 

The graph at the bottom of Figure~\ref{fig:ycsb_10rmw} shows
the same experiment under low contention. We find 
that locking once again outperforms the other concurrency control
protocols; however, the difference is much smaller. The reason why
locking still outperforms OCC
is that most OCC implementations (including our implementation of
Silo OCC \cite{tu13speedy})
requires threads to buffer their writes locally prior to making
writes visible in the database. Locking does not pay the overhead of
copying buffered writes to database records.
While OCC's write buffering is
similar to the multi-version systems's requirement of
creation of new versions, it has lower overhead because the same local
write buffer can be re-used by a single execution thread across many
different transactions (leading to better cache locality of the local
write buffers). In contrast, the multi-version systems need to write
\textit{different} locations on every update. 

Under low contention, the multi-version systems -- \systemName{},
Hekaton, and SI -- have similar performance. Hekaton and SI marginally
outperform \systemName{}, since our implementations of Hekaton and SI
do not include garbage collection and use array-based indices to
access records.

\subsubsection{2RMW-8R Workload}
\label{sec:ycsb_2rmw8r}

\begin{figure}
\centering
\includegraphics[clip,width=\linewidth]{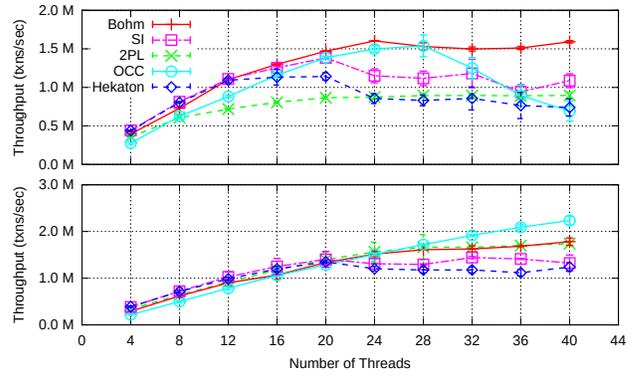}
\caption{YCSB 2RMW-8R throughput. Top: High Contention (Theta =
  0.9). Bottom: Low Contention (Theta = 0.0).}
\label{fig:ycsb_2rmw8r}
\end{figure}
This section compares \systemName{}'s throughput with each of our
baselines on a workload where each YCSB transaction performs two RMWs
and eight reads (2RMW-8R). In a high contention setting, we
expect that the multi-versioned systems will obtain
more concurrency than the single-versioned systems, since the reads
and writes of the same data items need not conflict under certain
circumstances. In particular, under SI, reads and writes never
conflict. Therefore, it is theoretically able to achieve more
concurrency than any of the other systems which guarantee
serializability and therefore have to restrict (to some degree), reads
and writes of the same data items in order to avoid the write-skew
anomaly (see Section~\ref{sec:mot}). In particular, \systemName{}
allows writes to block reads, but reads never block writes. In Hekaton,
reads also never block writes, but writes can cause transactions that
read the same data items to abort. In the single-version systems,
reads and writes \textit{always} conflict either via blocking (2PL) or aborting
(OCC).

The graph at the top of Figure~\ref{fig:ycsb_2rmw8r} shows the results
of this experiment under high contention. As expected, the
multi-versioned implementations outperform the single-versioned
implementations due to their ability to achieve higher concurrency,
and SI outperforms most of the other
systems due to the larger amount of concurrency possible when
serializable isolation is not required. 

Surprisingly, however, \systemName{} significantly outperforms 
SI.
We attribute this difference to aborts induced by write-write
conflicts in SI.
Under high contention this can lead to many
aborts and wasted work. Meanwhile, \systemName{} specifies the correct
ordering of writes to the same record across transactions in the
concurrency control layer, so that the transaction processing layer
simply needs to fill in placeholders and never needs to abort
transactions due to write-write conflicts
(Section~\ref{sec:txn_logic}). 
Like SI, Hekaton also
suffers from aborts and wasted work under high
contention. Interestingly, the Hekaton paper also implements a
pessimistic version of its concurrency control protocol, but finds
that it performs worse than the optimistic version, even under high
contention. This is because in the pessimistic version, reads
acquire read locks, and thus conflict with writes to the same record,
thereby reducing concurrency. Thus, a major contribution of
\systemName{} relative to Hekaton is a solution for allowing reads to
avoid blocking writes without resorting to optimistic mechanisms.

The graph at the bottom of Figure~\ref{fig:ycsb_2rmw8r} shows the
same experiment under low contention. OCC outperforms both
\systemName{} and locking as it employs a light-weight
concurrency control protocol, and does not
suffer from aborts under low contention. In particular, this workload
contains a significant number of reads, and read
validation is very cheap in Silo (which is our OCC implementation)
\cite{tu13speedy}. Note however, that \systemName{} is very close in
performance to OCC, despite the additional overhead of maintaining
multiple versions. 

The slope of the OCC, locking, and \systemName{} lines all decrease at higher thread
counts. We attribute this to the fact that our database tables span
multiple NUMA sockets. 

\begin{figure} \centering
\includegraphics[clip,width=\linewidth]{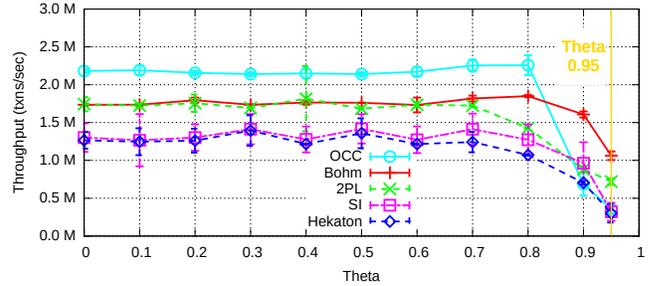}
\caption{YCSB 2RMW-8R throughput varying contention }
\label{fig:2rmw8r_varying}
\end{figure}

The most interesting part of the bottom of
Figure~\ref{fig:ycsb_2rmw8r} is the comparison of the three
multi-versioning implementations. With no contention, there are very
few aborts in optimistic schemes, nor any significant differences
between the amount of concurrency between the three
schemes. Therefore, one might expect all three implementations to
perform the same. However, we find that this is not the case; Hekaton
and SI are unable to scale beyond 20 cores.
We attribute Hekaton and SI's poor performance to contention on global
transaction timestamp counter. 
Hekaton and SI use a global 64-bit counter to assign
transactions their begin and end timestamps. In order to obtain a
timestamp, both systems atomically increment the value of the counter
using an atomic \textit{fetch-and-increment} instruction
(\texttt{xaddq} on our x86-64 machine). The counter is incremented at
least twice for \textit{every} transaction, regardless of the presence
of actual conflicts \footnote{The counter may be
incremented more than twice if a transaction needs to be re-executed
due to a concurrency control induced abort.}. At high thread counts,
SI and Hekaton are bottlenecked by contention on this global counter.
This observation is significant because it indicates that database
designs which rely on centralized contended data-structures are
\textit{fundamentally} unscalable. \systemName{}'s avoidance of this
prevalent limitation of multi-version concurrency control protocols
is thus an important contribution.

Figure~\ref{fig:2rmw8r_varying} further illustrates the fundamental
issue with Hekaton and SI's inability to scale. The graph shows the
throughput of each system (at 40 threads) while varying the degree of
contention in the workload. We use the same 2RMW-8R workload. The
graph indicates that both Hekaton and SI have identical performance
under low to medium contention, as they are both limited by the
timestamp counter bottleneck. Only under high contention does a new
bottleneck appear, and prevents the timestamp counter from being the primary
limitation of performance. 

\subsubsection{Impact of Long Read-only Transactions}
\label{sec:readonly_expt}

In this section, we measure the effect of long running read-only
transactions on each of our baselines. We run each baseline on a
workload consisting of a mix of update and read-only transactions.
Update transactions are the low contention 10RMW YCSB transactions from
Section~\ref{sec:ycsb_10rmw}.  Read-only transactions read 10,000
records -- chosen uniformly at random -- from the database. 

\begin{figure}
  \centering
  \includegraphics[clip,width=\linewidth]{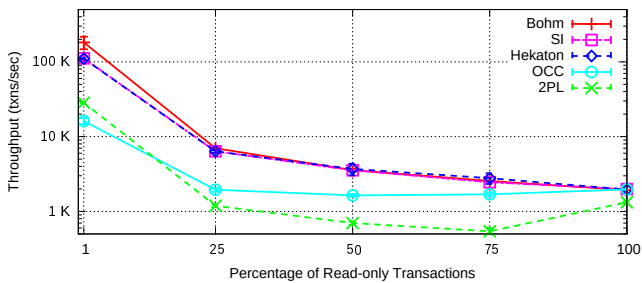}
  \caption{YCSB throughput with long running read-only transactions.}
  \label{fig:read_only_pct}
\end{figure}

Figure~\ref{fig:read_only_pct} 
plots the overall throughput of each system while varying the fraction
of read-only transactions in the workload. 
Figure~\ref{fig:read_only_1pct} provides a
detailed comparison between each system when 1\% of the transactions
are read-only.
When a small fraction of
the transactions are read-only (1\%), we find that the multi-version
systems outperform OCC and locking by about an order of magnitude.
This is because single-version systems cannot overlap the execution of read-only
transactions and update transactions. In the multi-versioned
systems, read-only transactions do not block the execution of
conflicting update transactions (and vice-versa) because read-only
transactions can perform their reads as of a timestamp which precedes
the earliest active update transaction.  
We also find that
\systemName{} significantly outperforms Hekaton and SI. 
We attribute
this difference to \systemName{}'s
read-set optimization (Section~\ref{sec:placeholders}), which ensures
that \systemName{} can obtain a reference to the version of a
record required by a transaction without accessing any
preceding or succeeding versions. In contrast, in Hekaton and
SI, if the version required by a transaction, $v_i$, has been
overwritten, then the system must traverse the list of succeeding
versions $v_n$, $v_{n-1}$, ..., $v_{i+1}$ (where $n > i$) in order to
obtain a reference to $v_i$. Version traversal overhead is not
specific to our implementations of Hekaton and SI -- it is inherent in
systems which determine the visibility of each transaction's writes
after the transaction has finished executing. Thus, version traversal
overhead is unavoidable in all conventional multi-version
systems. 

 \begin{figure}
 \centering
 \begin{tabular}{ |c|c|c| } 
   \hline
    & Throughput (txns/sec) & \% \systemName{}'s Throughput \\
   \hline
   \systemName{} & 181,565 & 100\% \\
   SI & 111,345 & 64.32\% \\
   Hekaton & 110,105 & 60.64\% \\
   2PL & 28,401 & 15.64\% \\
   OCC & 16,152 & 8.89\% \\
  \hline
 \end{tabular}
 \caption{YCSB throughput with 1\% long running read-only transactions.}
 \label{fig:read_only_1pct}
 \end{figure}

As the fraction of read-only transactions increases, the throughput of
each system drops. This is because each read-only transaction runs for
a significantly greater duration than update transactions (read-only
transactions read 10,000 records, while update transactions perform 
an RMW operation on 10 records). When the workload consists of 100\%
read-only transactions, all systems exhibit nearly identical
performance. This is because the workload does not contain
any read-write nor write-write conflicts. 

\subsection{SmallBank Benchmark}
\label{sec:small_bank}
Our final set of experiments evaluate \systemName{}'s performance on the
SmallBank benchmark \cite{cahill09serializable}. This benchmark
was used by Cahill et al.\ for their research on serializable
multi-versioned concurrency control.
SmallBank is designed to simulate a banking application.
The application consists of three tables, (1) Customer, a table which maps a
customer's name to a customer identifier, (2) Savings, a table whose rows
contain tuples of the form <Customer Identifier, Balance>, (3) Checking, a table
whose rows contain tuples of the form <Customer Identifier, Balance>.
The application consists
of five transactions:
(1) $Balance$, a read-only transaction which reads a single customer's checking
and savings balances,
(2) $Deposit$, makes a deposit into a
customer's checking account, 
(3) $TransactSaving$, makes a deposit or
withdrawal on a customer's savings account,
(4) $Amalgamate$, moves all funds from one
customer to another,
(5) $WriteCheck$, which writes a check against an account.
None of the transactions update the customer table --- only the Savings and
Checking tables are updated.

The number of rows in the $Savings$ and $Checking$ tables is equal to the number
of customers in the SmallBank database. We can therefore vary the degree of
contention in our experiments by changing the number of customers; decreasing
the number of customers increases the degree of contention in the SmallBank
workload. 

The transactions in the SmallBank workload are much smaller than the
transactions in the YCSB workload from the previous section. Every transaction
performs reads and writes on between 1 and 3 rows. Each record in the
$Savings$ and $Checking$ tables is 8 bytes long.
In comparison, our
configuration of the YCSB workload performs exactly 10 operations on each
transaction, and each record is of size 1,000 bytes. In order to make
the SmallBank transactions slightly less trivial in size, each
transaction spins for 50 microseconds (in addition to performing the
logic of the transaction).

\begin{figure}
\centering
\includegraphics[clip,width=\linewidth]{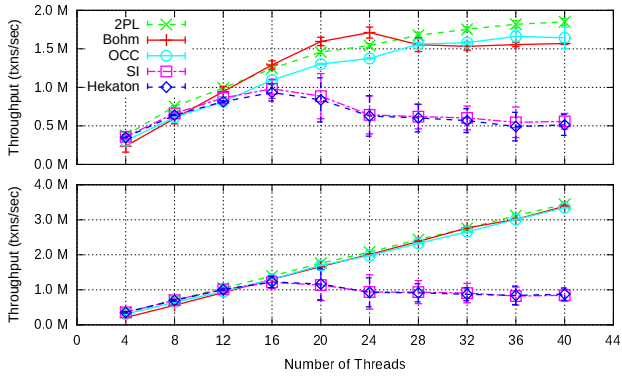}
\caption{Small Bank throughput. Top: High Contention (50
  Customers). Bottom: Low Contention (100,000 Customers).}
\label{fig:small_bank}
\end{figure}

Figure~\ref{fig:small_bank} shows the results of our experiment. 
The graph at the top of Figure~\ref{fig:small_bank} shows the results
under high contention (the number of SmallBank
customers is set to 50). Although locking once again performs best under high
contention, the difference between locking and \systemName{}
is not as large as in the contended 10RMW YCSB experiment
(Section~\ref{sec:ycsb_10rmw}). There are two reasons for this
difference:
 
First, as explained in Section~\ref{sec:ycsb_10rmw},
\systemName{} must pay the cost of bringing two \textit{different}
sets of memory words into cache on a read-modify-write operation, one
corresponding to the version that needs to be read, the second
corresponding to the version to be created. Since SmallBank's 8-byte
records are smaller than YCSB's 1000-byte records, the cost of
this extra memory access is smaller. As a consequence, the relative difference
between \systemName{} and locking is smaller.

Second, the workload from Section~\ref{sec:ycsb_10rmw} was 100\% RMW
transactions. In contrast, a small part of the SmallBank workload
(20\% of all transactions) consist of  read-only $Balance$ transactions.
Multi-versioned approaches such as \systemName{} are thus able to
increase the concurrency of these transactions, since reads do not
block writes.

Both Hekaton and SI's throughput drop under high contention due to
concurrency control induced aborts. 
At 40 threads, SI outperforms Hekaton by about
50,000 transactions per second because it suffers from fewer aborts
while validating transactions. Note that the abort-related drop in
performance of Hekaton and SI is greater than OCC. This is
because the contention on the timestamp counter for the
multi-versioned schemes (Hekaton and SI)
increases the time
required to get a timestamp. Since the SmallBank transactions are so
short, this time to acquire a timestamp is a nontrivial percentage of
overall transaction length. Hence, the transactions are effectively
longer for Hekaton and SI than they are for OCC, which leads to more
conflict during validation, and ultimately more aborts.

The graph at the bottom of Figure~\ref{fig:small_bank} shows the
results of the same experiment under low contention. We find that locking, OCC,
and \systemName{} have similar performance under this
configuration. As mentioned previously, the cost of RMW operations on
SmallBank's 8-byte records is much smaller than RMW operations on
YCSB's 1000-byte records. 

As we saw in previous experiments
(Section~\ref{sec:ycsb_2rmw8r}), we find that both Hekaton and SI are
bottlenecked by contention on the global timestamp counter. When using
40 threads, \systemName{} is able to achieve throughput in excess of 3 million
transactions per second, while Hekaton and SI achieve about 1 million
transactions per second; a difference of more than 3x. 

\section{Related Work}
\label{sec:rw}
\textbf{Concurrency Control Protocols.}
Timestamp ordering is a concurrency control technique in which the serialization
order of transactions is determined by assigning transactions monotonically
increasing timestamps
 \cite{bernstein81distributed,bernstein80sdd-1}.
In order to 
commit, conflicting transactions must execute in an order that is consistent 
with their timestamps. 
Reed designed a
multi-version concurrency control protocol based on timestamp ordering
 \cite{reed83implementing}.
Unlike single-version timestamp ordering,
reads are always successful, but readers may cause 
writers to abort and
the database needs to track the timestamp of each read in order to abort 
writers. In contrast, \systemName{}'s design guarantees that reads
never block writes and does require any kind of tracking when a
transaction 
reads the value of a record.
By eliminating writes to shared-memory, \systemName{}
greatly improves multi-core scalability.

Snapshot Isolation (SI) is a multi-version concurrency control protocol which
guarantees that transactions read database state as of a single
snapshot. SI guarantees that transactions that read the value of a
particular record never block transactions that write the record (and
vice-versa) by not detecting read-write conflicts among concurrent
transactions. SI is thus susceptible to serializability
violations. 
Unlike SI, \systemName{} guarantees serializability.
\systemName{} guarantees that reader transactions never block writer
transactions, but, unlike SI, \textit{does not} guarantee the converse.  
\\

\textbf{Multi-versioned Systems.}
Serializable Snapshot Isolation (SSI) is a modification to snapshot
isolation that guarantees serializability  \cite{cahill08serializable}.
SSI is based on the following fact about serializability violations in
the context of snapshot isolation:
if a serializability violation occurs, then the corresponding serialization
graph of
transactions contains two consecutive anti-dependency edges
 \cite{fekete05making}. 
SSI, therefore, tracks anti-dependencies among concurrent transactions 
on the fly and aborts transactions when it finds
a sequence of two consecutive
anti-dependency edges.
In contrast, \systemName{} ensures that the serialization graph
does not contain cycles \textit{prior} to actually executing transactions;
Consequently, \systemName{} 
does not need to track any anti-dependency edges at runtime.

Larson et al.\ propose techniques for optimistic and 
pessimistic multi-version concurrency control in the context of main-memory
databases \cite{larson11concurrency}. Their techniques address several
limitations of traditional systems. For instance, their optimistic validation 
technique does not require the use of a global critical section.
However, their design uses a
global counter to generate timestamps (that is accessible to many
different threads), and thus inherits the scalability
bottlenecks associated with contended global data-structures.
In contrast \systemName{} avoids the use of a global counter to
generate transactions' timestamps.

Neumann et al.\ propose an optimistic multi-version concurrency
control protocol that minimizes version maintenance overhead
\cite{neumann2015multiversion}. Their protocol allows transactions's
undo buffers to satisfy reads, minimizing the overhead of
multi-versioning relative to single-version systems.  However, their
protocol's scalability is bottlenecked by the use of a contended
global counter to generate transaction timestamps. Furthermore, like
Larson et al.\@'s optimistic protocol \cite{larson11concurrency},
their protocol aborts reading transactions in the presence of
read-write conflicts among concurrent transactions (Section
\ref{sec:guaranteeing}).
\\

\textbf{Multi-core Scalability.}
Silo is a database system designed for main-memory, multicore machines
\cite{tu13speedy}. Silo implements a variant of optimistic concurrency
control, and uses a decentralized timestamp based technique to
validate transactions at commit time.  \systemName{} shares some of
Silo's design principles. For instance, it uses a low contention
technique to generate timestamps to decide the relative ordering of
conflicting transactions. Unlike Silo, \systemName{} does not
use optimistic concurrency control, thus, it is able to perform
much better on high-contention workloads for which optimistic
concurrency control leads to many aborts.

Pandis et al.\ propose a data-oriented architecture (DORA) in order to eliminate
the impact of contended accesses to shared memory by transaction execution
threads \cite{pandis10data}. DORA partitions a database among
several physical cores of a multi-core system and
executes a disjoint subset of each transaction's logic on multiple
threads,
a form of intra-transaction parallelism. \systemName{}
uses intra-transaction parallelism to decide the \textit{order} in
which transactions must execute. However, the execution of a transaction's logic
occurs on a single thread. 

Jung et al.\ 
propose techniques for improving the scalability of 
lock-managers \cite{jung13scalable}. Their design includes the pervasive use of the
read-after-write pattern \cite{attiya11laws} in order to avoid repeatedly
``bouncing'' cache-lines due to cache-coherence
 \cite{anderson90performance,mellor91algorithms}. In addition, to avoid the cost
of reference counting locks, they use a
technique to \textit{lazily} de-allocate locks in batches.
\systemName{} similarly refrains from the use of reference counters to garbage
collect versions of records that are no longer visible to transactions.

Johnson et al.\ identified latch contention on high level intention
locks as a scalability bottleneck in multi-core databases
\cite{johnson09improving}. 
They proposed Speculative Lock Inheritance (SLI), a technique to reduce the
number of contended latch acquisitions.
SLI effectively \textit{amortizes} the cost of contended latch
acquisitions across a batch transactions by passing \textit{hot} locks
from transaction to transaction without requiring calls to the lock
manager. \systemName{} similarly amortizes synchronization across
batches of transactions in order to scale concurrency control.
\\

\textbf{Deterministic Systems.}
Calvin \cite{thomson12calvin} is a deterministic
database system that executes transactions according to a pre-defined total
order. Calvin uses deterministic transaction ordering to reduce the impact
of distributed transactions on scalability. Furthermore, Calvin uses a
modular architecture and
separates key parts of concurrency control from transaction execution
\cite{modular-calvin}. Although similar to
\systemName{} with its focus on scalability and modularity, Calvin is a
single-versioned system and uses locking to avoid read-write and
write-write conflicts, while \systemName{} is multi-versioned and
ensures that reads do not block writes. Furthermore, Calvin is focused
on horizontal shared-nothing scalability, while \systemName{} is
focused on multi-core scalability. 

Very lightweight locking (VLL) reduces lock-manager overhead by co-locating
concurrency control related meta-data with records
\cite{ren13lightweight}.  Unlike \systemName{}, VLL is not designed for systems with large
number of cores because every transaction must execute a global
critical section before it can execute.

H-Store \cite{hstore} uses a shared-nothing
architecture consisting of single-threaded partitions in order reduce the impact
of lock-manager overhead \cite{harizopoulos08glass}, and logging overhead
 \cite{malviya14rethinking}. However, performance degrades rapidly if
 a workload contains multi-partition transactions. Furthermore,
 sub-optimal performance is observed if some partitions have more work to
 do than others. \systemName{}
 achieves scalability without doing a hard-partitioning of the data ---
 it is thus less susceptible to skew problems and does not suffer from
 the multi-partition transaction problem. 
\\

\textbf{Dependency Graphs.}
Whitney et al.\ propose a deterministic concurrency control in which
transactions are executed according to a pre-defined total order
\cite{shasha-nocc}. Their system derives concurrency by constructing a
graph of transactions, which defines a partial order on transactions
based on their conflicts.
\systemName{} also pre-defines the
order in which transactions must execute, but its design is
fundamentally motivated by multi-core scalability. In contrast,
Whitney et al.~'s system contains several
centralized bottlenecks which inhibit multi-core scalability (e.g., the set of transactions that
are ready to execute is maintained in a centralized
data-structure).

Faleiro et al. describe a technique for \textit{lazily} evaluating transactions
in the context of deterministic database systems \cite{faleiro14lazy}.
This lazy database design
separates concurrency control from transaction execution --- a design
element that is shared by  
\systemName{}. However, \systemName{} does not
process transactions lazily, and is far more scalable due to its use
of intra-transaction parallelism, and avoiding writes to shared
memory on reads. Furthermore, \systemName{} is designed to be a generic
multi-versioned concurrency control technique, and is motivated by 
existing limitations in multi-version concurrency control systems.

\section{Conclusions}
\label{sec:concl}
Most multi-versioned database systems either do not guarantee
serializability or only do so at the expense of significant reductions
in read-write concurrency. In contrast \systemName{} is able to
achieve serializable concurrency control while still leveraging the
multiple versions to ensure that reads do not block writes. Our
experiments have shown that this enables \systemName{} to significantly
outperform other multi-versioned systems. Further, for workloads
where multi-versioning is particularly helpful (workloads containing a
mixture of reads and writes at high contention), \systemName{} is able
to outperform both single-versioned optimistic and pessimistic systems, without giving up
serializability. \systemName{} is the first multi-versioned
database system to accomplish this in main-memory multi-core
environments. 

{\bf Acknowledgments} This work was sponsored by the NSF under grant 
IIS-1249722 and by a Sloan Research Fellowship. We thank Phil
Bernstein, Alexander Thomson, and the anonymous VLDB 2015 reviewers for
their insightful feedback on earlier drafts of this paper.
\vspace{-.03cm}

\begin{scriptsize}
\bibliographystyle{abbrv}
\bibliography{ref}
\end{scriptsize}
\end{document}